\documentclass[twocolumn,showpacs,preprintnumbers,amsmath,amssymb,prl]{revtex4-2}

\usepackage{booktabs}
\usepackage{graphicx}
\usepackage[expansion=false]{microtype}
\usepackage{siunitx}
\usepackage{xcolor}
\definecolor{boxgray}  {HTML}{D8D8D8}
\definecolor{bordergray}{HTML}{707070}
\usepackage{tikz}
\usetikzlibrary{arrows.meta, calc}
\tikzset{
  solidarr/.style={-{Stealth[length=4pt,width=3pt]}, line width=0.8pt},
  dasharr/.style={-{Stealth[length=4pt,width=3pt]}, line width=0.6pt,
                  dashed, color=black!60},
  elabel/.style={fill=white, text=black, inner sep=1.5pt, font=\scriptsize},
}
\usepackage[colorlinks=true,
            linkcolor=blue!60!black,
            citecolor=blue!60!black]{hyperref}

\begin{document}

% ── Title and authors ─────────────────────────────────────────────────────────
\title{\textbf{Statistical Mechanics of Household Income and Wealth:\\
  Derivation from Firm Dynamics via Maximum Entropy\\
  and Mixture Aggregation}}

\author{Robert T.~Nachtrieb}
\affiliation{MIT Sloan School of Management, Cambridge, Massachusetts 02142, USA}
\date{\today}
% ── Abstract ─────────────────────────────────────────────────────────────────
\begin{abstract}
The distribution of income and wealth in developed economies exhibits a
robust two-class structure: an exponential (Boltzmann--Gibbs) bulk
covering $\sim\!97\%$ of the population, and a power-law (Pareto) tail
in the upper $\sim\!3\%$.  We derive this structure from first principles
via an explicit mechanistic chain: Gibrat's law for firm growth implies a
Zipf firm-size distribution; maximum entropy applied to within-firm wages,
combined with mixture aggregation across firms, yields a
Boltzmann--Gibbs income distribution with temperature $T_y$ for employees;
additive-noise wealth dynamics with a reflecting wall at zero produce
a Boltzmann--Gibbs employee wealth distribution with temperature $T_w$.
For firm owners, multiplicative capital returns produce a Pareto wealth
tail with exponent $\alpha_w = 1/\theta$, where $\theta$ encodes how total
returns scale with firm size.  The empirical value $\alpha_w \approx 1.30$
\cite{Yakovenko2009} is reproduced with no tuned parameters from the
observed firm value scaling $V = V_0(s/s_0)^{0.77}$~\cite{Axtell2001}, and
simultaneously yields the first quantitative estimate of the
returns-per-employee size exponent: $\zeta = \theta - 1 \approx -0.23$.
For empirical values $\nu \approx 0.3$, $c \approx 0.81$, $k \approx
0.15$ (BEA long-run savings rate $\approx 5\%$), the model gives
$T_w/T_y \approx 1.7\,\text{yr}$, i.e.\ lower-class households hold
roughly 1--2 years of income as wealth, with the precise ratio
depending on savings and tax rates and testable cross-country.
As a parameter-free empirical test, firms near zero profit have a
cash martingale whose first-passage time gives establishment exit rate
$\sim t^{-1/2}$; convolving with the Zipf firm-size distribution yields
firm-level exit rate $\sim t^{-1/2}\!\log t$, with apparent exponent
$b = 0.295 \pm 0.03$, confirmed against BDS firm-age data with no free parameters.
\end{abstract}
\maketitle

% ── I. Introduction ──────────────────────────────────────────────────────────
\section{Introduction}

The two-class structure of income and wealth distributions --- an
exponential (Boltzmann--Gibbs, GB) body and a Pareto tail --- has been
documented extensively by Yakovenko and co-workers
\cite{Yakovenko2009,DY2001} and is strikingly stable across time and
countries for the GB portion, while the Pareto tail fluctuates with
capital markets.  The historical growth of the Pareto tail and its
implications for inequality have been analyzed in depth by
Piketty~\cite{Piketty2014}.  Yet the \emph{mechanism} linking firm-level
microeconomics to household-level distributions has remained implicit in
the econophysics literature.

The closest antecedent is Aoki and Nirei \cite{AokiNirei2017}, who
jointly derive a Pareto income tail and Zipf firm-size distribution from a
neoclassical framework with optimisation assumed.  We provide a
complementary derivation using statistical mechanics and maximum entropy,
with no optimisation, and derive the BG distribution via a mixture
argument rather than assumed dynamics.  The resulting derivation yields a quantitative prediction for the ratio
of wealth to income temperature absent from purely entropy-based approaches,
and --- most strikingly --- reproduces the empirical Pareto wealth exponent
$\alpha_w \approx 1.30$ with no free parameters \emph{given} the Axtell
firm-value scaling $V = V_0(s/s_0)^{0.77}$~\cite{Axtell2001} and the
assumption that owner wealth tracks firm value, as summarised in Fig.~\ref{fig:1}.

% ── II. The Three-Sector Model ───────────────────────────────────────────────
\section{Three-Sector Economy}

We partition the economy into Firms, Households (subdivided into
Employees and Owners), and Government.  Solid arrows in
Fig.~\ref{fig:1}(a) denote flows that determine the distributions:
wages (Firms $\to$ Employees), consumption (Households $\to$ Firms), and
capital returns (Firms $\to$ Owners).  Dashed arrows denote fiscal flows
(taxes and government spending); we show below that the GB distributions
are robust to the precise form of fiscal flows, provided taxes and
transfers are additive in household income.

Money is conserved in real (inflation-adjusted) terms.  The boundary
conditions differ by sector: $w=0$ is \emph{reflecting} for households
(zero-wealth households remain in the system, receiving wages) and
\emph{absorbing} for firms (zero-cash firms exit, balanced by new
entrants at small size $s_0$).

% ── III. Firm Size Distribution ──────────────────────────────────────────────
\section{From Gibrat's Law to Zipf Firms}

Firm employment $s$ evolves by multiplicative (It\^{o}) noise:
\begin{equation}
  ds = \sqrt{2D_s}\,s\,d\mathcal{W},
  \label{eq:gibrat}
\end{equation}
where $\mathcal{W}$ is a Wiener process and $D_s$ [yr$^{-1}$] is the
diffusion coefficient in log-size space.  The zero-drift (Gibrat) assumption
\cite{Gibrat1931} is not imposed --- it \emph{emerges} from the Zipf
stationary condition: requiring $\partial_t p_f = 0$ in the
Fokker--Planck equation fixes the It\^{o} drift to zero via the
It\^{o}--Stratonovich correction term~\cite{SM}.
US Census Bureau Business Dynamics Statistics (BDS) \cite{BDS}
confirm the zero-drift prediction: employment expansion and contraction
rates are both flat at $r_{\rm churn} \approx 0.10\,\text{yr}^{-1}$ across all
firm sizes.  The Fokker--Planck equation with birth source
$\lambda\,\delta(s-s_0)$ and absorbing boundary at $s < s_0$ then has
the unique stationary solution
\begin{equation}
  S_f(s) \equiv P(S > s) \sim s^{-1}
  \quad (s \gg s_0),
  \label{eq:zipf}
\end{equation}
i.e.\ Zipf's law, confirmed empirically by Axtell \cite{Axtell2001} over
six decades of firm size.  A corollary: the size-weighted worker
distribution is \emph{flat} in $\ln s$ --- equal numbers of employees
work in each logarithmic interval of firm size.

\textit{Firm survival as empirical test.---}
Firms operating near zero profit ($\Pi/K \approx 0$) have a cash
reserve that follows a martingale; the time to bankruptcy is the
first-passage time of a Brownian motion, giving establishment exit
rate $\sim t^{-1/2}$ (L\'evy distribution).  A firm with $n$
establishments survives until the last exits, so
$T_{\rm firm} = \max(T_1,\ldots,T_n)$.  Averaging over the Zipf
size distribution ($\alpha = 1$) via the Laplace transform yields
$S_{\rm firm}(t) \sim t^{-1/2}\!\log t$, with apparent power-law
exponent $b = 0.295 \pm 0.03$ over a 30-year observation window.%
\footnote{The analytic power-law fit to the $t^{-1/2}\!\log t$ curve
over 1--30\,yr gives $b_{\rm app} \approx 0.32$; Monte Carlo simulation
at $\alpha = 1$ (Zipf) gives $b_{\rm sim} \approx 0.335$.  Both are
consistent with the BDS direct fit within the range spanned by the
empirical sector-level $\alpha$ values ($0.74$--$1.46$, bracketing Zipf);
see~\cite{SM}.}
This prediction, with no free parameters ($\alpha$ from CBP
establishment data, $b$ from BDS firm-age exit rates), is confirmed
in a companion paper~\cite{GP}.

% ── IV. From Zipf Firms to BG Income ─────────────────────────────────────────
\section{From Zipf Firms to Boltzmann--Gibbs Income: Mixture Argument}

\textit{Within-firm wage distribution.---}  Within a firm of size $s$
with total wage bill $Y_s$, the maximum entropy distribution subject to
non-negative wages and fixed mean $\bar{y}(s) = Y_s/s$ is exponential:
$p(y|s) = \bar{y}^{-1}e^{-y/\bar{y}(s)}$.  No assumptions about internal
hierarchy are required \cite{Jaynes1957}.

\textit{Mixture aggregation.---}  Empirically,
$\bar{y}(s) = y_0(s/s_0)^\epsilon$ with $\epsilon \approx 0.1$--$0.2$
\cite{BrownMedoff1989}: mean wages are nearly independent of firm size.
When $\epsilon \approx 0$, averaging the within-firm exponential over
the size-weighted Zipf distribution gives a mixture that is itself
approximately exponential:
\begin{equation}
  p(y) \approx \frac{1}{T_y}\,e^{-y/T_y}, \qquad T_y = y_0 = \langle y \rangle,
  \label{eq:income_bg}
\end{equation}
with corrections of order $\epsilon$ \cite{SM}.
This is a mixture argument, not the Central Limit Theorem:
the exponential form is preserved because the constraint (fixed mean wage)
is inherited by the aggregate, and maximum entropy applies at each level.
No Fokker--Planck equation governs employee income; the BG distribution
is a consequence of maximum entropy and mixture aggregation, not stochastic
dynamics.

% ── V. Household Wealth ───────────────────────────────────────────────────────
\section{Boltzmann--Gibbs Wealth and the Temperature Ratio}

\textit{Additive dynamics.---}  Household wealth $w$ (\$) evolves as
\begin{equation}
  dw = (y - e - \tau)\,dt + \sqrt{2D_w}\,d\mathcal{W},
  \label{eq:wealth_langevin}
\end{equation}
where $y$ is wage income, $e$ is consumption expenditure, $\tau$ is net
tax paid, and $D_w = (\sigma_y^2 + \sigma_e^2)\,\Delta t\,/\,2$ (\$$^2$\,yr$^{-1}$, with $\Delta t = 1$\,yr)
is the diffusivity from income and spending fluctuations.  Crucially, in the lower class all three terms $y$, $e$, $\tau$ are
\emph{additive}: set by labour market wages, consumption habits, and
tax schedules --- none depends on the household\'s current wealth level $w$.  The wealth Fokker--Planck equation with
reflecting boundary at $w=0$ has the stationary solution
\begin{equation}
  p_h(w) = \frac{1}{T_w}\,e^{-w/T_w},
  \qquad T_w = \frac{D_w}{|\mu_w|},
  \label{eq:wealth_bg}
\end{equation}
where $\mu_w = \langle y - e - \tau \rangle \approx 0$ is the small
residual drift (net savings after tax).  Taxes appear in the drift but,
being additive, do not alter the exponential form.

\textit{The temperature ratio.---}  Writing expenditure as $e = cy$ with marginal propensity to consume
$c$, and tax as fraction $k$ of income, the residual drift
is $|\mu_w| = (1-c-k)\,\langle y \rangle\,\text{yr}^{-1}$ and the
diffusivity is $D_w = \frac{1}{2}(\sigma_y^2+\sigma_e^2)\,\text{yr}$
where $\sigma_y^2 = \nu^2 T_y^2$ ($\nu \approx 0.3$ the year-over-year
income volatility relative to the mean) and $\sigma_e^2 = c^2\sigma_y^2$
since $e = cy$.
The natural unit of time is one year, so all quantities are in
[USD\,yr$^{-1}$] for flows and [USD] for stocks:
\begin{equation}
  \boxed{\frac{T_w}{T_y} = \frac{\nu^2(1+c^2)}{2(1-c-k)}\,\Delta t
  \approx 1.7\,\text{yr} \quad (\nu=0.3,\ c=0.81,\ k=0.15).}
  \label{eq:ratio}
\end{equation}
Both sides have units of years: $T_w$ [\$] divided by $T_y$
[\$/yr] gives [yr], and the right-hand side carries $\Delta t = 1$\,yr
explicitly.  Here $c \approx 0.81$ follows from the BEA long-run personal
savings rate $s \approx 5\%$ on disposable income via $c = 1-k-s(1-k)$.
The result --- lower-class households hold roughly 1--2 years of income
as wealth at steady state --- is testable cross-country via household
balance-sheet surveys (e.g.\ the Federal Reserve Survey of Consumer Finances).

\textit{Pareto tail.---}
Owner wealth scales with firm size as
$w = V_0\,(s/s_0)^\theta$,
where $V_0$~[\$] is the reference firm value at reference
size $s_0$ [workers], and $\theta \approx 0.77$ is the
firm-value size-scaling exponent
($V = V_0(s/s_0)^\theta$, from Axtell~\cite{Axtell2001}).
Since $w$ is a power-law function of the Zipf variable $s$,
the owner wealth survival function inherits a Pareto tail:
$S_{\rm own}(w) \sim w^{-\alpha_w}$ with $\alpha_w = 1/\theta \approx 1.30$.
For $\theta=1$ (wealth linear in size) $\alpha_w = 1$ (Zipf);
$\theta < 1$ gives $\alpha_w > 1$, consistent with the observed
upper-tail exponent~\cite{Yakovenko2009}.
Diversification raises $\alpha_w$ toward $\infty$ (index-fund limit);
the systematic treatment is in Gabaix~\cite{Gabaix2009}.

% ── VI. Discussion ────────────────────────────────────────────────────────────
\section{Discussion}

\begin{figure*}[t]
  \centering
  \resizebox{\textwidth}{!}{\input{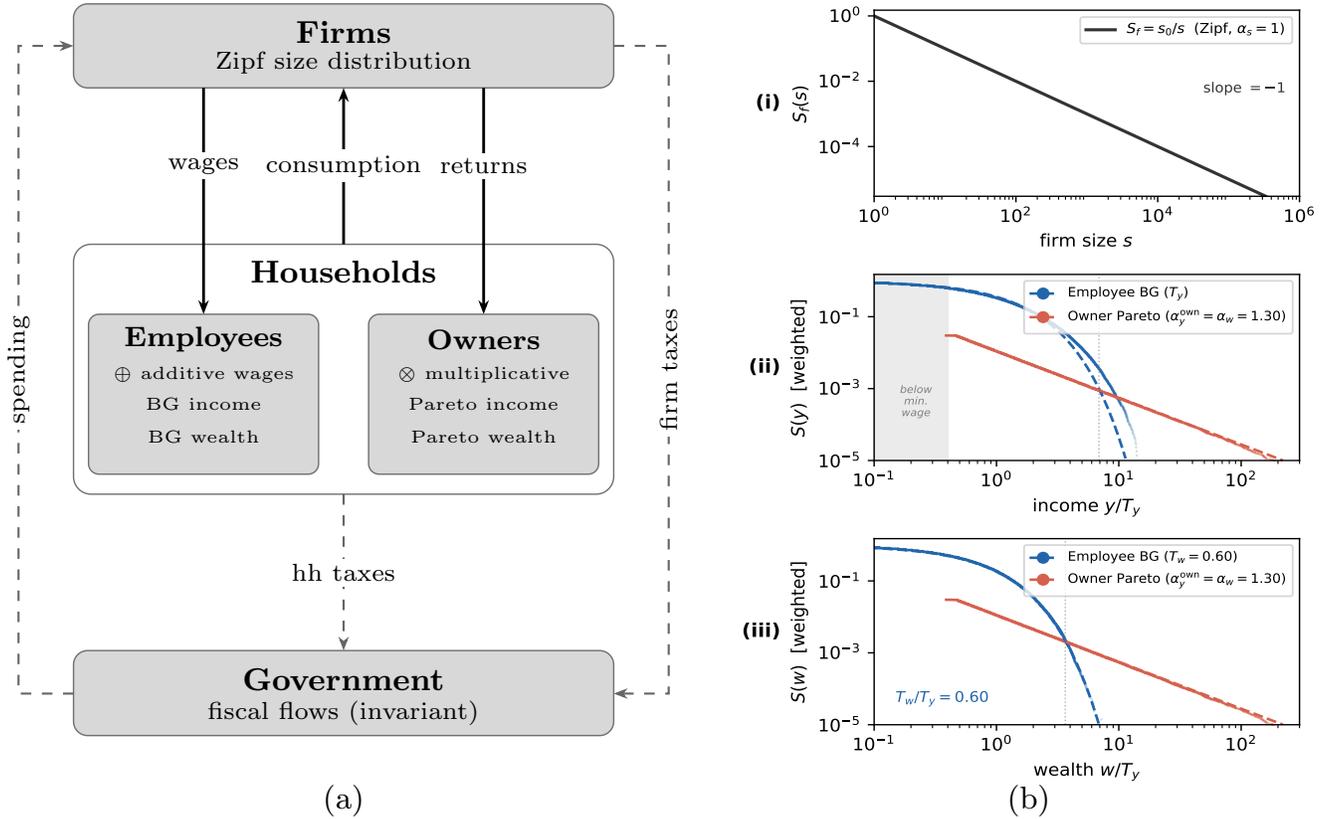}}
  \caption{(a) Three-sector money flow diagram.  Solid arrows: flows
    determining the distributions (wages, consumption, capital returns).
    Dashed arrows: fiscal flows (taxes, government spending), shown to
    leave the GB form robust.  Employees ($\oplus$ additive wages) yield
    BG income and wealth; Owners ($\otimes$ multiplicative returns) yield
    Pareto income and wealth.
    (b) Survival functions $S(x) = P(X > x)$ on log--log axes, from
    Monte Carlo simulation (dots) and analytic fits (dashed lines),
    population-weighted ($f_{\rm emp}=0.97$, $f_{\rm own}=0.03$).
    Firm size (i) follows Zipf ($S_f \sim s^{-1}$, slope $-1$).
    Income (ii) and wealth (iii) each show a BG bulk (concave, set by
    temperature $T_y$ or $T_w$) transitioning to a Pareto tail (straight
    line, slope $-\alpha_w$); the gray dotted vertical marks the crossover
    $x_1$ where the two populations contribute equally~[Eq.~(\ref{eq:x1})].
    The ratio $T_w/T_y \approx 1.7\,\text{yr}$ [Eq.~(\ref{eq:ratio})]
    is derived in the text; see~\cite{SM} for the relationship between
    $\alpha_w$ and the empirical income Pareto exponent.}
  \label{fig:1}
\end{figure*}

The causal tree implied by Fig.~\ref{fig:1}(a) has a main branch for
employees --- Gibrat diffusion $\to$ Zipf firms $\to$ max-entropy wages
$\to$ mixture aggregation $\to$ BG employee income $\to$ BG employee wealth --- and a
fork for owners: Zipf firms $\to$ Pareto owner wealth $\to$ Pareto owner
income~\cite{Gabaix2009,Benhabib2011,SM}.  A companion paper~\cite{GP}
derives the profit imperative $\eta^* \equiv (w/\kappa + 1/\tau)/(1-f_p)$
from BEA accounting identities and shows that sectors near
$\eta \approx \eta^*$ (zero profit) provide the cash martingales
underlying the firm survival prediction above.
The asymmetry between the two
branches is physically meaningful: employees earn income first and
accumulate wealth through savings; owners hold wealth first and collect
income as returns on that wealth.  Each step invokes the weakest
assumption sufficient to derive the result.

The key novelty is the mixture aggregation step: the GB income distribution
emerges from combining maximum entropy within firms with the approximate
constancy of mean wages across firm sizes ($\epsilon \approx 0$).
This is why the GB bulk is empirically stable --- it follows from
maximum entropy plus a mild empirical regularity --- while the Pareto tail
(driven by multiplicative capital returns) is volatile.

The Pareto exponent $\alpha_w = 1/\theta$ connects two independently measured
quantities with no tuned parameters.  Axtell~\cite{Axtell2001} reports
firm value (market capitalisation) scaling as $V = V_0(s/s_0)^{0.77}$; under
the assumption that owner wealth tracks firm value, $\theta = 0.77$ predicts
$\alpha_w \approx 1.30$, in quantitative agreement with the upper-tail
income Pareto exponent of Yakovenko and Rosser~\cite{Yakovenko2009}.
(The income and wealth Pareto exponents coincide in the limit of
proportional returns, which holds to leading order for large owner
wealth~\cite{SM}.)  Running the logic in reverse
--- which is arguably more striking --- both measurements together yield
the first quantitative estimate of the returns-per-employee size exponent:
$\zeta = \theta - 1 \approx -0.23$, implying larger firms generate
somewhat less profit per employee --- a new empirical target accessible
from IRS Statistics of Income or Census LEHD microdata~\cite{SM}.
It has been suggested~\cite{Gabaix_pc} that Axtell's $\theta = 0.77$
may understate the full-population value, since small listed firms
carry high growth expectations that inflate market cap relative to
employment.
To assess this, we estimated $\theta$ from Compustat (1990--2023,
$n > 300{,}000$ firm-years) using three dimensionally consistent
proxies for firm value: market capitalisation, book equity, and
enterprise value.
All three give $\theta \in [0.69, 0.76]$, corroborating Axtell
and inconsistent with $\theta = 1$~\cite{SM}.
Compustat covers listed firms only; the full-population $\theta$
for private firms remains unmeasured.
If it is closer to $1$, then $\alpha_w \to 1$ (Zipf wealth) and
$\zeta \to 0$ --- both testable with IRS or LEHD microdata.

The GB distributions are robust to a wide class of fiscal rules because
additive taxes and transfers shift $T_w$ quantitatively but do not alter
the exponential form.  For empirical values $\nu \approx 0.3$,
$c \approx 0.81$, $k \approx 0.15$, Eq.~(\ref{eq:ratio}) gives
$T_w/T_y \approx 1.7\,\text{yr}$, consistent with lower-class net worth
in the Federal Reserve Survey of Consumer Finances.  The precise ratio
is cross-country testable: economies with higher savings rates
($c \to 0$) or tax rates ($k \to 1-c$) should exhibit smaller $T_w/T_y$.

The BG--Pareto crossover point $x_1$ --- visible in
Fig.~\ref{fig:1}(b) as the kink in the survival function --- is
commonly treated as a fitting parameter~\cite{Yakovenko2009}.  In our
model it is an output, determined by the condition $S_{\rm BG}(x_1) =
S_{\rm Pareto}(x_1)$:
\begin{align}
  \frac{x_1}{T_y} + \alpha_w\ln\frac{x_1}{y_{\rm min}}
  = \ln\frac{f_{\rm emp}}{f_{\rm own}},
  \label{eq:x1}
\end{align}
where $f_{\rm emp}/f_{\rm own}$ is the ratio of employee to owner
\emph{population counts}.  For the US ($f_{\rm emp}/f_{\rm own}
\approx 97/3$), Eq.~(\ref{eq:x1}) gives $x_1 \approx 7\,T_y$
in income --- the structural boundary between the two classes,
set by demography rather than by distributional fitting.

The stationary BG wealth distribution holds when the residual drift
$\mu_w \approx 0$; for a growing economy ($\mu_w > 0$, $\sim 2\%$\,yr$^{-1}$)
the distribution acquires a correction of order $\mu_w T_w / D_w \ll 1$
for the empirical parameter range, leaving the exponential form robust.
The full dynamical treatment --- equations of motion for capital deepening,
the profit imperative, and the connection to firm survival --- is developed
in a companion paper~\cite{GP}.
Future work includes disaggregating Firms into consumer and contractor
sub-sectors and modelling retirement as an absorbing household state.
Supplemental Material \cite{SM} provides detailed derivations,
equilibrium and stability analysis of the distributions, finite-difference
integration of the coupled Fokker--Planck equations, and Monte Carlo
agent simulation.

% ── Figures ───────────────────────────────────────────────────────────────────

% ── Acknowledgements ──────────────────────────────────────────────────────────
\begin{acknowledgments}
The author thanks V.~Yakovenko for correspondence on the
Boltzmann--Gibbs income distribution, X.~Gabaix for feedback
on the Pareto wealth tail, and D.~Peterson for the insight
connecting capital productivity to embedded technology.
\end{acknowledgments}

% ── References ────────────────────────────────────────────────────────────────

\end{document}